\newif\ifapj
\apjtrue

\ifapj
  \documentclass[twocolumn]{aastex6}
\else
  \documentclass[preprint]{aastex6}
\fi

\usepackage{amsmath,apjfonts,amsfonts,amssymb,enumerate,graphicx,
  longtable,float,times,textcomp,verbatim,url,nicefrac}
\def\apj{ApJ}%
\def\apjl{ApJ}%
\def\apjs{ApJS}%
\def\ao{Appl.~Opt.}%
\def\aap{A\&A}%
\def\mnras{MNRAS}%
\def\pasp{PASP}%
\def\ssr{Space~Sci.~Rev.}%
\newcommand{\oiii}{[O{\sc iii}]\,$\lambda5007$}

\begin{document}

\title{A New Approach to the Internal Calibration of
  Reverberation Mapping Spectra}

\author{M.~M.~Fausnaugh\altaffilmark{1}}
\altaffiltext{1}{Department of Astronomy, The Ohio State University,
  140 W 18th Ave, Columbus, OH 43210, USA}

\begin{abstract}
  We present a new procedure for the internal (night-to-night)
  calibration of time series spectra, with specific applications to
  optical AGN reverberation mapping data.  The traditional calibration
  technique assumes that the narrow \oiii\ emission line profile is
  constant in time; given a reference \oiii\ line profile, nightly
  spectra are aligned by fitting for a wavelength shift, a flux
  rescaling factor, and a change in the spectroscopic resolution.  We
  propose the following modifications to this procedure: 1) we
  stipulate a constant spectral resolution for the final calibrated
  spectra, 2) we employ a more flexible model for changes in the
  spectral resolution, and 3) we use a Bayesian modeling framework to
  assess uncertainties in the calibration. In a test
  case using data for MCG+08-11-011, these modifications result in
  a calibration precision of $\sim\! 1$ millimagnitude, which is approximately a factor of
  five improvement over the traditional technique.  At this level,
  other systematic issues (e.g., the nightly sensitivity functions and
  Fe{\sc ii} contamination) limit the final precision of the observed
  light curves.  We implement this procedure as a {\tt python} package
  ({\tt mapspec}), which we make available to the community.
\end{abstract}

\section{Introduction}

Reverberation mapping \citep[RM,][]{Blandford1982, Peterson1993,
  Peterson2014} is a very successful way of exploring the spatially
unresolved structures in active galactic nuclei (AGN).  The
conspicuous broad emission lines observed in Seyfert 1 and quasar
spectra respond to continuum variations on weekly to monthly time
scales.  Measurements of the time delay between the continuum signal
and the emission line ``echoes'' establish the characteristic size of
the line-emitting gas.  This technique has become a primary means of
estimating the masses of super-massive black holes that are associated
with AGN activity \citep{Peterson2004, Bentz2015}.  On shorter or
longer time scales (less than a few days or greater than a month),
lags between the UV, optical, and IR continua provide a means of
applying RM to the accretion disk or the ``dusty torus'' (e.g.,
\citealt{Kishimoto2007, Shappee2014, Vazquez2015, Edelson2015,
  Fausnaugh2016}).  On even longer times scales (several years to
decades), narrow emission line reverberations can probe structures up
to several tens of parsecs across \citep{Peterson2013}.

Crucial to RM measurements is a precise estimate of the intrinsic
variability of the AGN.  Such estimates require a treatment of
extrinsic sources of variability, such as those created by differences
in observing conditions from night to night.  Studies that fail to do
this will attribute extrinsic variability to the intrinsic AGN
emission.  \citet{Barth2016} analyze an example of such a problem in
detail.  The usual approach is to model and remove these extrinsic
variations by assuming that some component of the AGN spectrum is
constant over the full time series.  The narrow \oiii\ emission line
serves as a practical choice, since it originates in an extended
region of the AGN (tens to hundreds of light years across) and should
be constant over the course of a typical RM campaign (a few months).
It is also relatively uncontaminated with other spectral
  features, although blending with variable Fe{\sc ii} emission and
  the red wing of H$\beta$ can sometimes be problematic.

The traditional implementation of the rescaling model is that of
\citet[hereinafter GW92]{vanGroningen1992}.  The GW92 model uses
  an empirical template to correct a series of observations for
differences in wavelength solution, attenuation, and spectral
resolution, and routinely reaches night-to-night precisions of 3--5\%.
As RM data have improved (e.g., \citealt{Denney2010, Grier2012,
  Du2016}), it has been possible to reach precisions closer to 1--3\%,
and sometimes even better (0.5--0.7\%, \citealt{Barth2015}).  Over the
last 25 years, only minor modifications have been applied to the
original GW92 approach.  For example, \citet{Barth2015} updated the
GW92 optimization procedure from a grid-search to a down-hill simplex
algorithm.  Occasionally, studies bypass the rescaling procedure all
together; \citet{Kaspi2000} and \citet{Du2014} corrected all
  extrinsic variations using simultaneous observations of comparison
  stars.  Another approach is to forgo the empirical template and
  model the narrow line emission with parametric functions.  This is
  the approach adopted by the Sloan Digital Sky Survey Reverberation
  Mapping project \citep{Shen2015,Shen2016} using the {\tt PrepSpec}
  software developed by Keith Horne.  \citet{Hu2016} recently employed
  a similar modeling technique to improve the calibration of RM data
  taken in 2008 of MCG-6-30-15 from 2\% to 0.5\%.

Considering the gains in computing resources over the last two decades
and the rise of alternative model-fitting techniques, we decided to
investigate more substantial modifications to the GW92 procedure.  In
\S2 we review the main elements of the GW92 rescaling model and
propose three improvements.  We then discuss a new model and fitting
procedure to implement these modifications, and we make our
implementation available to the community as a {\tt python} package
called {\tt mapspec} ({\bf M}CMC {\bf A}lgorithm for {\bf P}arameters
of {\bf
  Spec}tra).\footnote{\url{https://github.com/mmfausnaugh/mapspec}} In
\S3, we assess our method by applying it to new RM data for
MCG+08-11-011 (UCG 3374) that has been presented more completely
elsewhere (\citealt{Fausnaugh2017b}).  We find that the precision of
the night-to-night calibration increases by roughly a factor of five
using our new approach, and the final light-curve uncertainties are
dominated by intrinsic systematic effects that require more
complicated methods to address.  In \S4, we summarize these results,
and we include a brief appendix that discusses the influence of
correlated errors on our results.

\section{The Approach of GW92 and Proposed Improvements}

The GW92 approach aligns the \oiii\ line profile of some observed
spectrum $O$ to a reference spectrum $\hat R$ by applying a wavelength
shift, a flux rescaling factor, and a smoothing kernel.  The shift
accounts for differences in the wavelength solution, the rescaling
factor for differences in attenuation (e.g., atmospheric extinction),
and the smoothing kernel for differences in resolution (e.g., changes
in seeing or spectrograph focus).

To fit for these parameters, GW92 use a grid search.  The current
model parameters are applied to $O$ to create a rescaled spectrum
$\tilde O$, and the alignment with $\hat R$ is determined using the
difference spectrum $\tilde D = \hat R - \tilde O$. Near a narrow
emission line with constant flux (such as \oiii), the only difference
between $\hat R$ and $\tilde O$ should be intrinsic continuum and
broad-line flux variations.  Therefore, $\tilde D$ should be a smooth
function of wavelength, and GW92 use the $\chi^2$ of a low-order
polynomial fit to $\tilde D$ to measure of the alignment between $\hat
R$ and $\tilde O$:
\begin{align}
  \chi^2 = \sum_{\lambda} \frac{\left[ \tilde D(\lambda) - P(\lambda)\right]^2}{\sigma_{R}^2(\lambda) + \tilde \sigma_{O}^2(\lambda)}
\label{equ:gwchi}
\end{align}
where $P(\lambda)$ is the polynomial, $\sigma_R$ is the uncertainty on
$\hat R$\, and $\tilde \sigma_O$ is the uncertainty on $\tilde O$.
Note the $P$ is fit directly to $\tilde D$ and that $\chi^2$ is always
minimized with respect to the polynomial; the rescaling model is
considered optimized when $\chi^2$ is minimized with respect to
$\tilde D$ by finding the model parameters that best align $\tilde O$
with $\hat R$.

To account for changes in resolution, the GW92 model uses a Gaussian
smoothing kernel.  The main effect of changing the resolution is to
change the width of the observed emission lines.  GW92 therefore
parameterize the spectral resolution using the full width at half
maximum (FWHM) of the \oiii\ line.  Seeing variations are usually
small from night to night, so changes in resolution are generally
small compared to the width of narrow emission lines, and practically
negligible compared to the width of broad emission lines.

A complication arises because $\hat R$ may have a higher resolution
(lower FWHM) than $O$.  Since deconvolution is numerically unstable,
GW92 also test models where $\hat R$ is smoothed to match $O$---if the
resulting $\chi^2$ is smaller than smoothing $O$ to match $\hat R$,
then $O$ is inferred to have a lower resolution (higher FWHM) than
$\hat R$.  In these cases, the final rescaled spectrum $\tilde O$ is
not corrected for resolution, so as to avoid deconvolution.

This method has been very effective in past RM campaigns (e.g.,s
\citealt{Peterson2004, Bentz2009b, Denney2010, Barth2011, Grier2012,
  Barth2013, Pei2014, Barth2015}).  However, we suggest several
modifications to improve this approach.

\begin{enumerate}
\item {\bf Resolution.} Because GW92 chose to ignore resolution
  corrections for cases where $O$ is observed at lower resolution
  (greater FWHM) than $\hat R$, the final set of rescaled spectra
  exist at a variety of resolutions.  The minimum FWHM is defined by
  $\hat R$, since the model will smooth $O$ to match $\hat R$ whenever
  possible, while the maximum FWHM is set by the epoch with the worst
  resolution.  This means that any quantity calculated from the
  ensemble spectra (for example, the mean spectrum), has additional
  scatter from the heterogeneous resolutions.  We instead construct
  $\hat R$ so that its resolution matches the \emph{worst} resolution
  (largest FWHM) of the time series.  This will guarantee that the
  model always prefers to smooth $O$ to match $\hat R$, and $\hat R$
  will therefore define a single resolution of the rescaled spectra.

\item {\bf Smoothing Kernel.} While a Gaussian smoothing kernel is a
  good first-order approximation for changes in the resolution from
  night to night, the true kernel can be significantly more
  complicated.  For example, there are changes due to miscentering in
  the slit, changing spectrograph focus, guiding errors, and flexure
  in the telescope/optics system.  Using a smoothing kernel that is
  more complex than a Gaussian should lead to an improved nightly
  calibration, and we use Gauss-Hermite polynomials to parameterize
  this complexity.  A similar approach is often taken when measuring
  the line-of-sight velocity dispersion in galaxy spectra (e.g.,
  \citealt{ vanderMarel1993}).

\item {\bf Model Uncertainties.} While minimizing $\chi^2$ on a grid
  is computationally efficient, this approach can make estimating
  model uncertainties difficult.  GW92 did not rigorously consider
  uncertainties in the model parameters, although they show that these
  uncertainties are smaller than any ``by eye'' rescaling approach.
  We instead use a Bayesian framework when optimizing the rescaling
  parameters, which can naturally account for model uncertainties from
  the parameters' posterior probability distributions.
\end{enumerate}
\subsection{Reference Spectrum}
The first step of the calibration is to construct a high
signal-to-noise ratio (S/N) reference spectrum.  The reference defines
the flux scale of the calibrated time series, so the usual practice is
to combine spectra from photometric nights---this ensures that the
final time series corresponds to physical fluxes.  We construct the
reference by averaging the photometric spectra, weighting each
observation at epoch $t_i$ by the measurement uncertainty
$\sigma_O(\lambda)$.  The uncertainty in the reference is therefore
$\sigma_{R}(\lambda) = \left(\sum_i 1/\sigma_{O}^2(\lambda,t_i)
\right)^{-1/2}$.  Even with a moderate number of spectra, the
uncertainty in any observed spectrum will be much larger than the
uncertainty in the reference.

Our first modification relates to the resolution of the reference.
When averaging the photometric spectra, the resulting resolution
roughly corresponds to the average resolution of the input spectra (in
practice, we have found this to be the case, see \S3).  We then smooth
the reference so that its resolution matches the lowest resolution of
all the observations.  Like GW92, we estimate the spectral resolution
with the width of the narrow \oiii\ emission line, using the FWHM of
Gaussians fit to the observed line profiles.  Although the true \oiii\
line profiles are more complex, this simple method provides a good
relative comparison.  The \oiii\ FWHM in the smoothed reference then
defines the final resolution of the entire time series.

\subsection{Model Details and Fitting Procedure}
We model the rescaled spectrum $\tilde{O}$ from an observed spectrum
$O$ as
\begin{align}
\tilde{O}(\lambda) = a\int_{-\infty}^{\infty} O(\lambda - s)K(\lambda -  \lambda')\,d \lambda'\label{equ:model}
\end{align}
where $a$ is a flux scaling factor, $s$ is a wavelength shift, and $K$
is a smoothing kernel.  We assume that $O$ has an accurate relative
flux calibration (that the slope of the continuum is correct) and that
$O$ is free of aperture effects (that the flux from extended sources
such as the host-galaxy and narrow line region is fixed; see
\citealt{Peterson1995}).  The smoothing kernel is a sum of
Gauss-Hermite polynomials
\begin{align}
K(\lambda - \lambda ') = e^{u^2/2} \sum_{i=0}^N b_i H_i(u) & &;& &
u = \frac{\lambda - \lambda '}{w}\label{equ:kernel}
\end{align}
where the lowest order term $i=0$ is a simple Gaussian with width $w$,
$H_i(u)$ are the Gauss-Hermite polynomials following the definition
of \citealt{vanderMarel1993}, and $b_i$ are coefficients to be
optimized.  We adopt
\begin{align}
b_0 = 1 & &
b_1 = b_2 = 0
\end{align}
and truncate the series at $N=4$, which gives the coefficients a
simple interpretation: $b_3$ quantifies asymmetric deviations from a
Gaussian, similar to skewness, and $b_4$ quantifies symmetric
deviations, similar to kurtosis \citep{vanderMarel1993}.  Since the
$H_i$ are orthogonal on the Gaussian weighting function, $b_3$ and
$b_4$ are uncorrelated, simplifying the fitting procedure.  Under this
formalism, we can recover a simple Gaussian model by taking $N=0$, or
we can add arbitrary complexity by extending the series beyond $N=4$.

We simultaneously fit for $a$, $s$, $w$, and any $b_i$ using Markov
Chain Monte Carlo (MCMC) methods.  The MCMC searches for parameters
that minimize
\begin{align}
  \chi^2 = \sum_\lambda \frac{\left[\hat{R}(\lambda) - \tilde{O}(\lambda)\right]^2}{\sigma_R^2(\lambda) +
    \tilde{\sigma}_{O}^2(\lambda)}\label{equ:loglikely}
\end{align}
where $\hat{R}(\lambda)$ is the reference spectrum,
$\sigma_R(\lambda)$ is its uncertainty, and
$\tilde{\sigma}_{O}(\lambda)$ is determined by standard error
propagation on the observed spectrum:
\begin{align}
  \tilde{\sigma}_{O}^2(\lambda) =  a^2\int_{-\infty}^{\infty} \sigma_{O}^2(\lambda - s) K^2(\lambda -
  \lambda ')\,d\lambda ' \label{equ:obserror}.
\end{align}
Here, $\sigma_{O}^2(\lambda -s)$ is the uncertainty from
interpolation.  Linear interpolation is usually adequate with
$\sigma_{0}^2 (\lambda_{i+1} -s) = (1 - x)^2\sigma_{0}^2
(\lambda_{i+1}) + x^2\sigma_{0} ^2(\lambda_{i})$ where $x$ is the
fractional pixel shift $s/(\lambda_{i+1} - \lambda_{i})$.

Minimizing $\chi^2$ is equivalent to maximizing the log-likelihood of
the data given the model, assuming normally distributed and
uncorrelated residuals.  We discuss the possible influence of
correlated residuals in the Appendix.  Equation \ref{equ:loglikely}
also implicitly assumes uniform (uninformative) priors on the
parameters,
but alternative choices of prior parameter distributions are fully
supported by the {\tt mapspec} implementation (see below).

Only non-variable parts of $\hat R$ and $\tilde{O}$ should be
compared, i.e., the narrow \oiii\ line.  We isolate the emission line
by subtracting a local linear-interpolation of the underlying
continuum.  We remove large wavelength offsets of the spectra by
cross-correlation, which finds the shift to the nearest pixel, so that
the parameter $s$ should usually be less than a pixel.  When
performing the fit, we ignore 5\% of the data on each end of the
fitting window---this avoids edge effects from the convolution and
helps stabilize the number of degrees of freedom (the overlap of the
spectra can change for large values of the wavelength shift $s$).  The
kernel width $w$ is restricted to be greater than or equal to half a
pixel---at smaller values, the smoothing kernel is indistinguishable
from a $\delta$-function.
We also explicitly normalize $K$ so that $\int_{-\infty}^{\infty}
K\,d\lambda = 1$, which conserves flux and helps prevent correlations
between the rescaling factor $a$ and parameters that define $K$.
Finally, we restrict $b_3$ and $b_4$ to lie between $-0.3$ and $0.3$.
Empirical experiments indicate that this range is a small enough for
the MCMC chain to converge quickly, while it is large enough to
produce a wide range of line profiles.

The MCMC procedure automatically produces posterior probability
distributions for each parameter.  Examination of the posterior
distributions allows us to quickly assess the quality of the fit.  For
example, observations taken in bad conditions will have poorly
constrained parameters because of low S/N at each pixel.  Such a
diagnostic is not available using the traditional GW92 grid-search.

We have developed a {\tt python} package that implements the above
model and fitting procedure, which we call {\tt mapspec} ({\bf M}CMC
{\bf A}lgorithm for {\bf P}arameters of {\bf Spec}tra) and make freely
available.  The software is object-oriented to facilitate modularity
and extensibility. The package also includes data structures that
naturally organize spectroscopic data and provide useful operations
and analysis methods, such as interpolation, rebinning, line
extraction/integration, velocity percentile calculation, etc.  These
data structures are well suited for use on any spectrum with emission
lines.  A variety of data formats are supported, including ascii,
  comma-separated value, and fits files.  Template scripts are
provided that help construct a reference spectrum and fit the model to
a given time series.  Finally, the software supports a flexible
implementation of priors during the fitting procedure---the user may
specify any analytic probability density function as a prior on any
parameter in the model.  Posteriors distributions from previous {\tt
  mapspec} runs can also be imported for use as priors on future fits.

\section{Test Case: MCG+08-11-011}
We compared our new approach to that of GW92 using the time series
spectra of MCG+08-11-011 (UGC 3774) from a recent RM campaign
(\citealt{Fausnaugh2017b}).  Data were taken on the 1.3m McGraw-Hill
telescope at the MDM observatory in Spring of 2014.  MCG+08-11-011 is
a low-redshift source ($z\approx 0.02$, {\it V} $\approx 14.8$ magnitude)
that exhibited strong and coherent variability during the RM campaign.

\subsection{Reference Spectrum}
\begin{figure}
\includegraphics[width=0.5\textwidth]{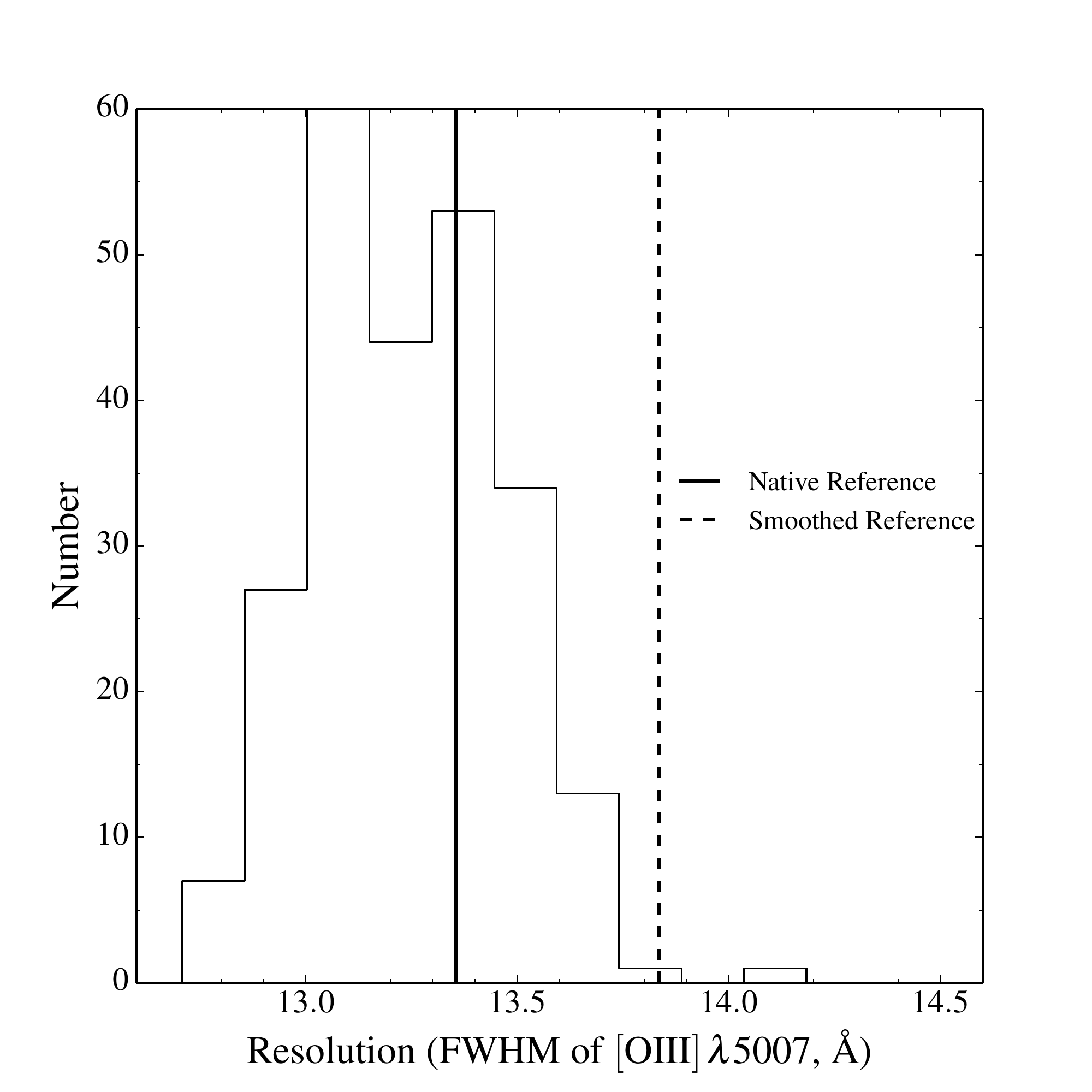}
\caption{Distribution of spectral resolutions for the full time
  series, estimated by the FWHM of Gaussians fit to the narrow \oiii\
  line profiles. The resolution of the reference spectrum constructed
  from photometric nights is shown with the solid line, the resolution
  of the reference after smoothing is shown with the dashed
  line.\label{fig:refres}}
\end{figure}

We constructed the reference from 18 spectra taken on six photometric
nights as reported by the observers.  We checked the photometric list
of observations by calculating the \oiii\ line flux in each spectrum.
We subtracted a local linearly-interpolated continuum underneath the
line and integrated the remaining flux using Simpson's method.  We
then applied iterative $3\sigma$ clipping to the array of line flux
measurements, which removed three spectra from the photometric list.
Our final estimate of the \oiii\ line flux is $(6.13\pm 0.02)\times
10^{-13}{\rm \ erg\ cm^{-2}\ s^{-1}}$, calculated from the remaining
15 spectra.

We combined these spectra with a noise-weighted average to make an
initial estimate of the reference spectrum at its ``native''
resolution.  During this step, we used MCMC methods to fit for
  the wavelength shift that best aligns the \oiii\ profiles of the
  spectra in a least-squares sense.
  The reference sets the final wavelength grid, so it is important to
  adopt a single and consistent wavelength solution when averaging.
  However, the absolute accuracy of this solution does not matter for
  the purpose of the night-to-night calibration (in practice, we chose
  a wavelength solution accurate to 0.5\,\AA, as measured by night-sky
  lines).

We estimated the resolutions of the input spectra with the FWHM of
Gaussians fitted to \oiii\ line profiles.  The mean FWHM of these
spectra is is 13.35\,\AA, very close to the value measured in the
weighted average spectrum of 13.36\,\AA.\footnote{A more rigorous
  estimate of the spectral resolution is obtained by subtracting in
  quadrature the intrinsic line width from the observed line
  width. \citet{Whittle1992} gives the intrinsic FWHM of the \oiii\
  line in MCG+08-11-011 as 605 km s$^{-1}$.  This corresponds to
  10.52\,\AA\ in the observed frame and gives a spectral resolution of
  8.22\,\AA.  However, the only effect of this correction is to shift
  the distribution in Figure \ref{fig:refres} by a constant amount, so
  we omit this correction here.  Note that \citet{Whittle1992} measure
  the FWHM directly from the line profile instead of using a Gaussian
  fit.  If we measure the FWHM in the ``native'' reference in the same
  way, we find a value of 12.63\,\AA.  This implies a spectral
  resolution of 6.99\,\AA, the value adopted by Fausnaugh et al. in
  preparation.}  Next we created a smoothed reference whose resolution
matches the lowest resolution spectrum of the full time series.  The
distribution of \oiii\ FWHM measurements from the full time series is
shown in Figure\,\ref{fig:refres}.  The FWHM measurement above
14.0\,\AA\ appears to be an outlier that could be explained by
especially bad seeing or large guiding errors that caused the target
to move in the slit.  We therefore took the next-largest FWHM of the
distribution (13.75\,\AA) as the worst resolution of the time series.
We smoothed the ``native'' reference with a Gaussian kernel of ${\rm
  FWHM} = 3.26$\,\AA\ to produce a smoothed reference with a
resolution of ${\rm FWHM} = 13.84$\,\AA.  This will set the final
resolution of the rescaled time series.

\subsection{Comparison of {\tt mapspec} and GW92}

\begin{figure}
\includegraphics[width=0.5\textwidth]{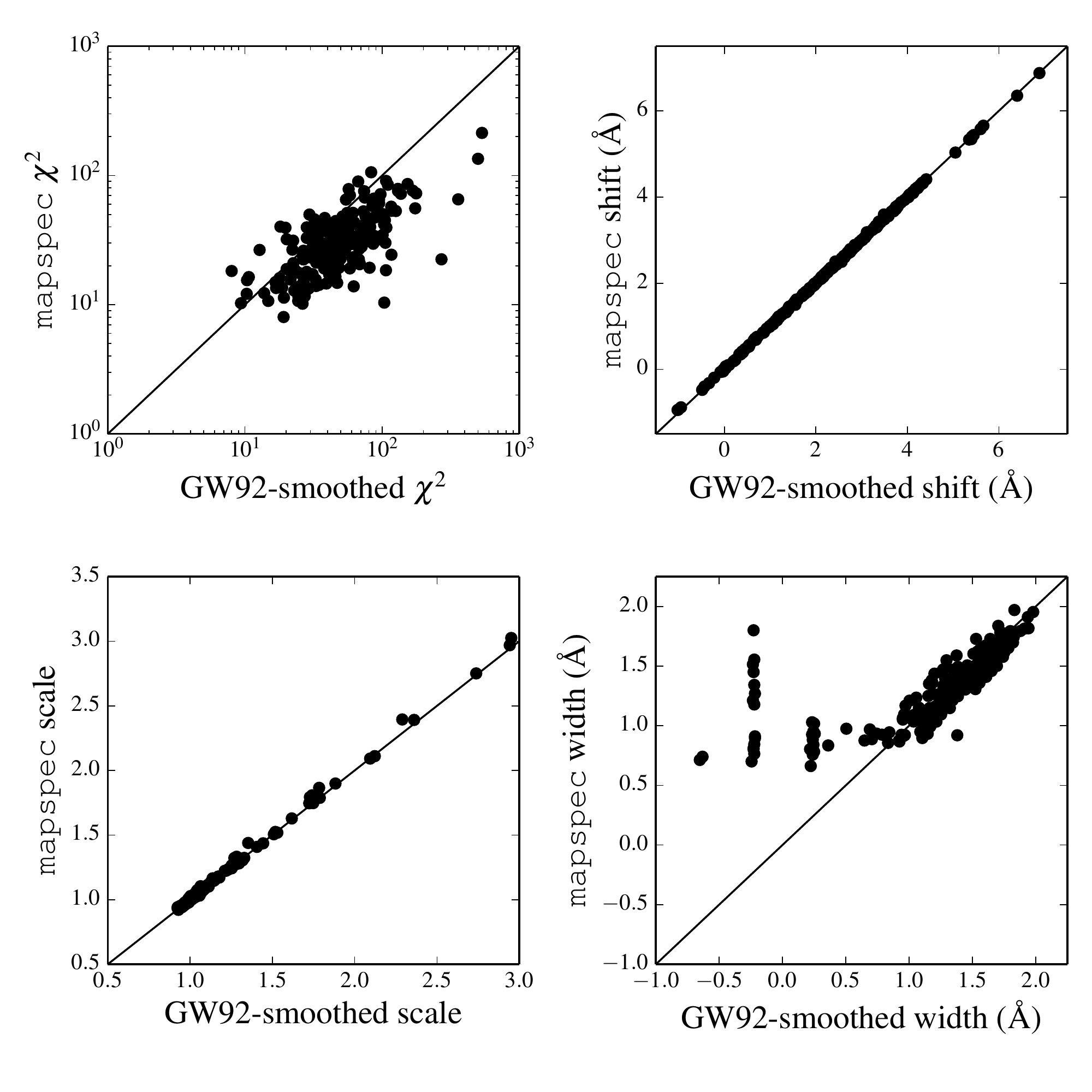}
\caption{Comparison of parameters fit by the GW92 algorithm and the
  {\tt mapspec} procedure.  The solid lines show one-to-one relations.
  A negative width indicates that the GW92 method prefers to smooth the
  reference to match the observation, and the output spectra are not
  smoothed in these cases. \label{fig:params}}
\end{figure}

\begin{figure}
\includegraphics[width=0.5\textwidth]{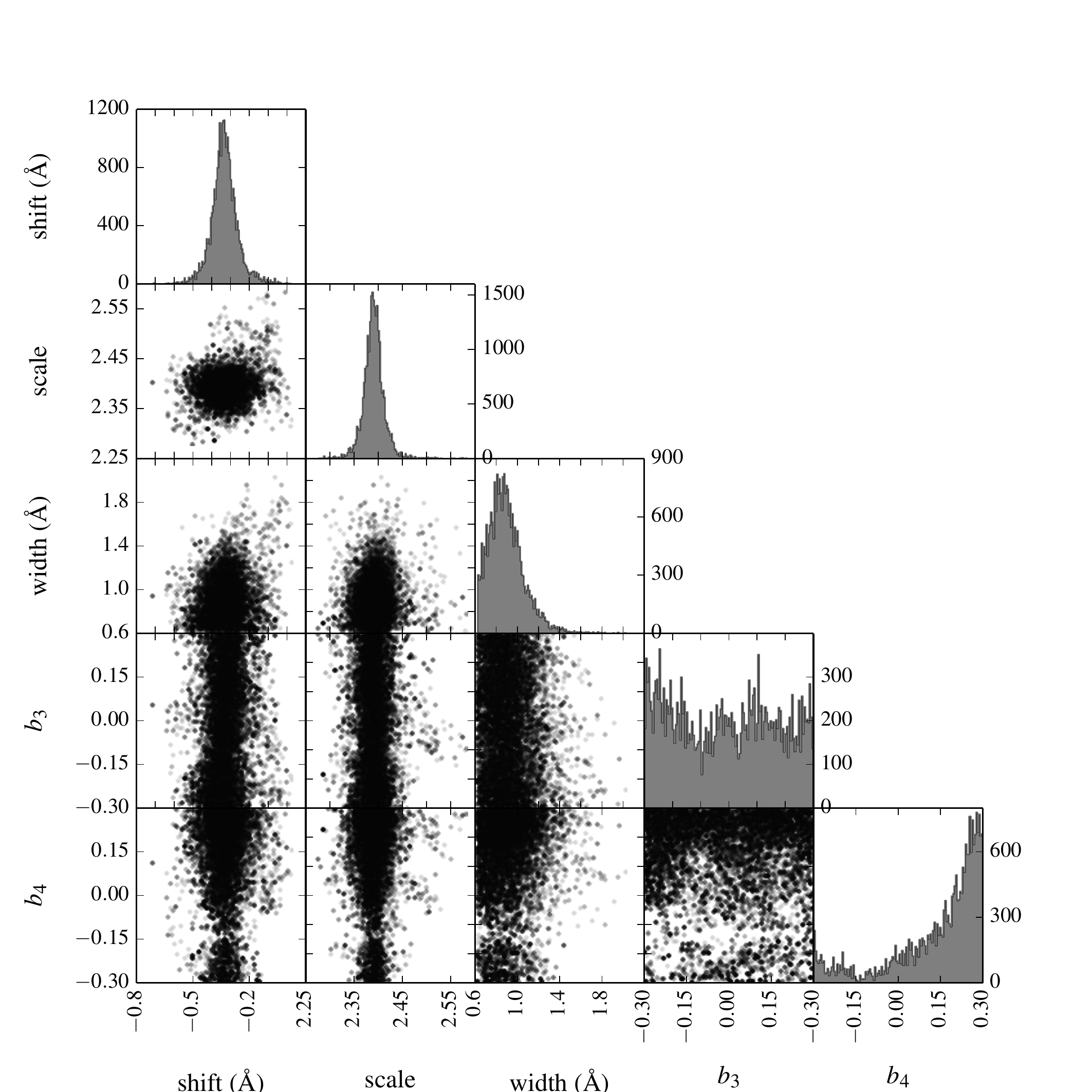}
\caption{Example of posterior parameter distributions for the
  observation on 2014 April 17 UT.  The conditions were partly cloudy
  with persistent thick cirrus, and the spectrum requires significant
  rescaling.  The model prefers a peakier line profile, but the
  skewness is unconstrained (the best fit value of $b_3$ is
  0.0).  \label{fig:posterior_example}}
\end{figure}

We compared the {\tt mapspec} fits to two different implementations of
the GW92 method.  The first implementation rescales the time series to
match the ``native'' reference and allows a heterogeneous set of
output resolutions, as per the original GW92 approach.  The second
method aligns the time series to the smoothed reference with the same
GW92 model.  This allows us to independently compare the effects of
adopting a single resolution for the final spectra with the effects of
the {\tt mapspec} model and fitting procedure.  We designate the first
approach the GW92 method and the second approach the GW92-smoothed
method.

In Figure \ref{fig:params}, we compare the parameters found by {\tt
  mapspec} to those of the GW92 scaling procedure (for the smoothed
reference only, since we do not expect the kernel widths to match for
different reference spectra).  There is excellent agreement between
the two models, especially for the shift and rescaling parameters.
The dispersion is somewhat larger for the width parameter, but our
models are not identical and we do not expect a perfect match.
Negative widths for the GW92 procedure indicate spectra with FWHM
greater than that of the reference, and are not smoothed when
rescaling.  In principle, the algorithm should always prefer to smooth
the observation to match the reference; the cases with negative FWHM
might be due to the grid-search algorithm, which can converge to a
local rather than global minimum, or to limitations in the smoothing
model itself (a pure Gaussian). The values of $\chi^2$ for both fits
also track each other reasonably well, although they are not defined
in the same way (see Equations \ref{equ:gwchi} and
\ref{equ:loglikely}).  The {\tt mapspec} model tends to have smaller
$\chi^2$ values than the GW92 procedure, which may be a result of the
additional {\tt mapspec} parameters.

In terms of performance, it takes 8352 seconds of user-time (2.3
hours) to calibrate 240 observations of MCG+08-11-011 (on average,
three observations every night for 80 nights).  For comparison, the
GW92 grid-search requires 75 seconds to run on the same data.
Although {\tt mapspec} takes over 100 times longer to run than the
GW92 algorithm, it provides the additional benefits of a more general
and flexible model, and saves the posterior probability distributions
of the parameters.  Since each spectrum is considered independently,
the amount of user time can be greatly reduced simply by distributing
the calibration processes across several computers.

In Figure \ref{fig:posterior_example}, we show a set of posterior
distributions for one spectrum taken on 2014 April 17 UT.  Conditions
were partly cloudy with persistent cirrus clouds, and the observation
required significant rescaling.  The best-fit parameters are $a =
2.38$, $s = -0.34$\,\AA, $w = 0.92$\,\AA, $b_3 = 0.00$, and $b_4 =
0.29$.  The wavelength shift $s$ and flux rescaling factor $a$ are
very well constrained, with the sizes of the central 68\% confidence
intervals equal to 0.01\,\AA\ and 1\% of the of the median rescaling
factor, respectively.  The posterior distribution of the kernel width
$w$ is somewhat broader, and while there is no constraint on the
skewness, the model clearly prefers a peaky line profile.

In general, we find that the wavelength shift and flux rescaling
parameters are similarly well-constrained for all observations, while
the kernel shapes show a wide combination of widths, skewness, and
kurtosis.  In one case (2014 January 29 UT), we found that the scaling
parameter was poorly constrained (the central 68\% interval of the
posterior distribution was 6\% of the median value).  Visual
inspection of this spectrum showed an anomalous ``shelf'' on the blue
wing of the \oiii\ line that is not present in any other spectrum.
For this night, weather conditions consisted of patchy clouds, and the
anomalous feature may be due to movement of the target in the slit if
the guide star was temporarily lost during the observation.  We
therefore exclude this observation from the final data set.  This
procedure demonstrates the diagnostic utility of carefully examining
the posterior probability distributions.

One caveat is that there is very little Fe{\sc ii} emission in
  MCG+08-11-011, and the \oiii\ line is not blended with H$\beta$.  In
  other AGN, the \oiii\ line is strongly blended with these variable
  components and it is not easy to isolate the narrow line flux with a
  local linearly-interpolated continuum.  We have tested our method on
  objects farther along the Eigenvector 1 sequence \citep{Boroson1992}
  with weak \oiii\ and very strong Fe{\sc ii} emission, and we found
  that neither {\tt mapspec} nor the GW92 approach provided a
  reasonable calibration.  The {\tt mapspec} procedure may still be
  useful for such spectra, but a more sophisticated means of
  separating the narrow line from the continuum, Fe{\sc ii}, and
  possibly H$\beta$ emission would be required.

\subsection{Mean and RMS Spectra}
\begin{figure*}
\includegraphics[width=\textwidth]{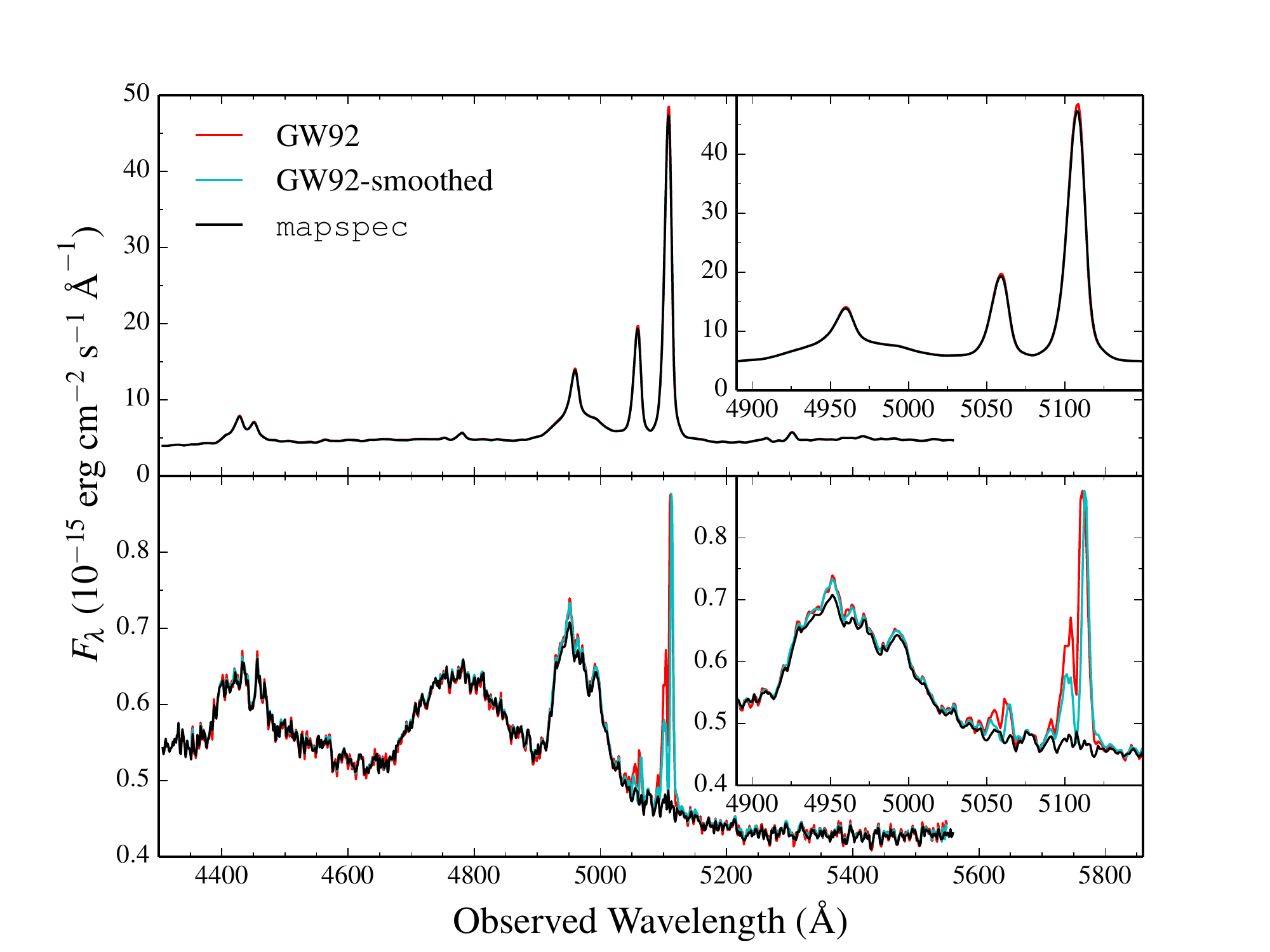}
\caption{Comparison of the mean and rms spectra derived from the three
  rescaling methods.  GW92 is the original \citet{vanGroningen1992}
  method, using a reference constructed from photometric nights and
  left at its native resolution.  GW92-smoothed uses the same
  procedure but with a smoothed reference (see \S2.1). {\tt mapspec}
  is the result from the model described in \S2.2.  Note the dramatic
  suppression of the [O{\sc iii}] residuals in the {\tt mapspec}
  approach.\label{fig:spec}}
\end{figure*}

We compare the output spectra from different rescaling procedures
using the mean and root-mean-square (rms) residual spectra of the time
series.  The mean spectrum $\bar F(\lambda)$ is the average of the
rescaled spectra, weighted by the measurement uncertainties after
propagation through the model.  The rms spectrum is
\begin{align}
F_{\rm rms}(\lambda) = \sqrt{ \frac{1}{N-1}\sum_i^N \left[ \tilde O(\lambda,t_i) - \bar F(\lambda)\right]^2}.
\end{align}
Figure \ref{fig:spec} shows the mean and rms spectra from the GW92,
GW92-smoothed, and {\tt mapspec} methods.  The mean spectra are almost
indistinguishable---the only visible difference is a peakier \oiii\ in
the original GW92 method, which is expected because the reference was
not smoothed for this procedure.  The rms spectra are also very
similar---the continuum and broad lines are virtually
indistinguishable between the three methods.

There are noticeable differences in the rms spectrum near the [O{\sc
  iii}]\,$\lambda 4959$ and \oiii\ line profiles, which are
highlighted in the insets in Figure \ref{fig:spec}.  We see large
residuals across both narrow line profiles for the GW92 method.  The
GW92-smoothed method has similar but smaller residuals, mainly
confined to the wings of the lines.  This shows that a substantial
fraction of the rms residuals are due to the heterogeneous resolution
of the original GW92 approach.  These [O{\sc iii}] residuals vanish
when using the {\tt mapspec} model, presumably due to the more
flexible line profiles afforded by the Gauss-Hermite polynomials.
Close inspection of the H$\beta$ line-profile also shows that the
narrow component has been more cleanly removed from the rms spectrum
by the {\tt mapspec} approach.

Suppression of the [O{\sc iii}] and H$\beta$ narrow line residuals is
an important benefit of the {\tt mapspec} approach.  The rms spectrum
isolates the variable part of the spectrum, so the velocity of the
reverberating gas is usually measured from the line-width of the rms
line-profile.  The presence of spurious variability, such as residuals
from narrow line profiles, can affect these line-width measurements.
This effect is not very important in these data, since the broad
H$\beta$ line has essentially no overlap with the [O{\sc iii}] lines,
and the differences between the GW92 and {\tt mapspec} H$\beta$ rms
profiles are small.  However, in other AGN the red wing of H$\beta$ is
often blended with the [O{\sc iii}] lines, and/or the narrow H$\beta$
component is more significant compared to the broad component (see,
e.g., \citealt{Grier2012} and \citealt{Fausnaugh2017b}, for
several examples).  For such objects, suppressing systematic errors in
the nightly calibration represents a more substantial improvement in
the reliability of line-width measurements.  

Finally, suppressing the [O{\sc iii}] residuals may provide a
  means of identifying Fe{\sc ii} variability.  The Fe{\sc ii}
  multiplet lines at 4924\,\AA\ and 5018\,\AA\ are blended with the
  narrow [O{\sc iii}] lines, and these features are impossible to
  measure underneath the large [O{\sc iii}] residuals caused by the
  GW92 approach.  Using {\tt mapspec}, the smoothness of the continuum
  at these wavelengths suggests that it may be possible to detect
  Fe{\sc ii} variability for other objects. However, we know that very
  strong Fe{\sc ii} emission will cause the calibration to fail
  (\S3.2), so more testing is required to assess {\tt mapspec}'s
  potential in this regard.

\subsection{Light Curves}
\begin{figure*}
\includegraphics[width=\textwidth]{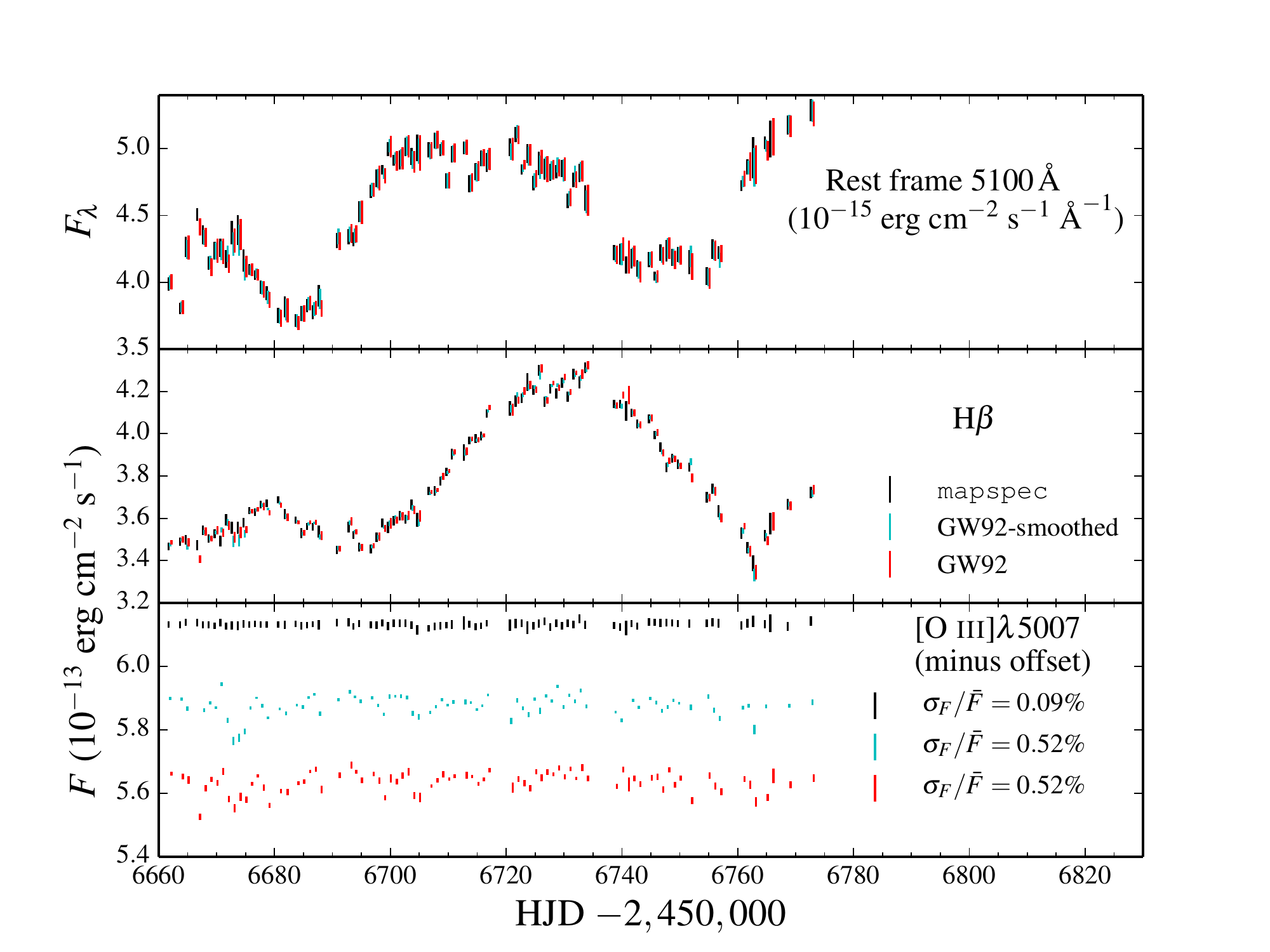}
\caption{Comparison of light curves extracted from spectra calibrated
  with the three different procedures.  GW92, GW92-smoothed, and {\tt
    mapspec} correspond to the same methods as in Figure
  \ref{fig:spec}.  For clarity, the data from different rescaling
  methods have been given a small offset along the abscissas, as well
  as along the ordinates for the \oiii\ light curves.  The choice of
  rescaling method changes the amount of scatter in the \oiii\ line
  light curve by a factor of five, quantified by the fractional
  root-mean-square scatter $\sigma_{F}/\bar F$.\label{fig:lc}}
\end{figure*}

The final goal of the night-to-night calibration is to measure
intrinsic flux variations, i.e., light curves. In Figure \ref{fig:lc},
we show three light curves extracted from the time series
spectra---the continuum at 5100\,\AA\ (rest-frame), the integrated
broad H$\beta$ line, and the integrated \oiii\ line.  The continuum
light curve was estimated using the average flux density over the
observed wavelength region between 5190 and 5230 \AA, and its
uncertainty is the sample standard deviation in this window.  The
H$\beta$ line flux was estimated by subtracting a local linear
approximation of the underlying continuum and integrating the
remaining flux using Simpson's method.  We did not correct for the
narrow H$\beta$ component, which contributes a constant flux-offset to
these measurements.  Line flux uncertainties were estimated using a
Monte Carlo approach: the spectrum was perturbed $10^3$ times by
random Gaussian deviates scaled to the flux uncertainty at each
wavelength (including correlations introduced by smoothing, see the
Appendix), and the line was re-extracted and integrated for each
iteration.  We adopt the width of the central 68\% confidence interval
of the resulting distribution as an estimate of the line-flux
uncertainty.  An identical procedure was applied to the \oiii\ line.

It is clear that the choice of calibration procedure makes only a
small difference for the continuum and H$\beta$ light curves.  This
result might be expected based on the similarity of the mean and rms
spectra at these wavelengths (Figure \ref{fig:spec}).  However, the
{\tt mapspec} approach greatly reduces the scatter in the \oiii\ line
light curve compared to the GW92 method.  Following \citet{Barth2013},
the scatter in the \oiii\ light curve serves as an estimate of the
remaining uncertainty in the night-to-night calibrations.  The
fractional scatter in the {\tt mapspec} \oiii\ light curve is 0.09\%
(or zero if adjusted for the line-flux uncertainties, see
\citealt{Barth2013, Barth2015}).  This is roughly a factor of five
gain over the 0.52\% scatter of the original GW92 approach.  For other
data sets, precisions of 3\% to 5\% are sometimes the best attainable
with the GW92 approach, in which case a factor of five improvement is
much more meaningful. 	From the same 2014 campaign, we also have
  data for 3C\,382 and Mrk\,374.  The GW92 approach resulted in
  calibrations precise to 1.83\% for 3C\,382 and 1.95\% for Mrk\,374,
  measured using the \oiii\ light curves in the same way as for
  MCG+08-11-011.  These are more typical results, and the {\tt
    mapspec} approach improves the calibrations to 0.92\% for 3C\,382
  (a factor of 2) and 0.62\% for Mrk\,374 (a factor of 3).

\begin{figure}
\includegraphics[width=0.5\textwidth]{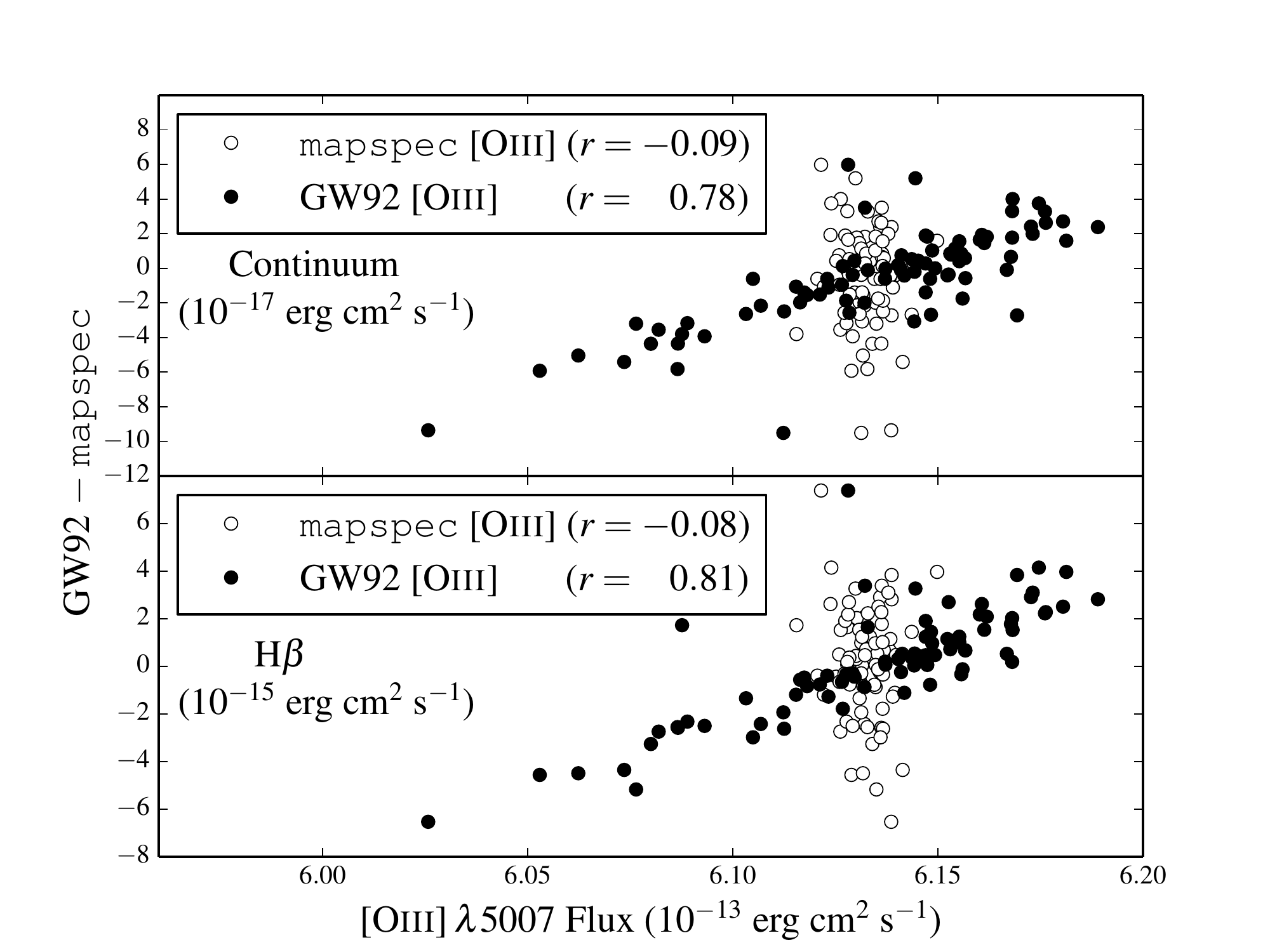}
\caption{Differences between the continuum and H$\beta$ light curves
  from different calibration procedures (red points minus black points
  in Figure \ref{fig:lc}) as a function of \oiii\ line flux.  \oiii\
  line fluxes from the GW92 procedure are shown with solid points,
  \oiii\ line fluxes from {\tt mapspec} are shown with open points.
  The Pearson $r$ correlation coefficients are shown in the
  legends. These correlations show that errors in the GW92 \oiii\
  calibration introduce noise in the continuum and line light
  curves.\label{fig:lc_corr}}
\end{figure}

While it is not obvious from Figure \ref{fig:lc}, the differences
between the {\tt mapspec} and GW92 H$\beta$ and continuum light curves
are largely due to the GW92 calibration errors.  In Figure
\ref{fig:lc_corr}, we show the light curve differences (the red points
minus the black points in Figure \ref{fig:lc}) as a function of \oiii\
flux from both procedures.  Since the {\tt
  mapspec} \oiii\ light curve is virtually flat, we see no correlation
of the H$\beta$ and continuum light curve differences with the {\tt
  mapspec} \oiii\ fluxes.  However, the light curve differences are
strongly correlated with the GW92 \oiii\ fluxes, illustrating how the
GW92 calibration errors add noise to the resulting H$\beta$ and
continuum light curves.

The {\tt mapspec} calibration precision is nominally $\sim\!  0.1$\%
(millimagnitues).  However, there are other systematic effects that
limit the final precision of the light curves:

\begin{itemize}
\item Errors in the relative flux calibrations of the spectra will
  affect the observations regardless of any rescaling model.  The
  relative flux calibration depends on the nightly sensitivity
  functions, which are themselves calculated from observations of
  standard stars.  The night-to-night repeatability of the sensitivity
  functions therefore depends on the choice of standard star, the
  observing conditions, and even the choice of image reduction and
  fitting techniques.  It is likely that this uncertainty
  enters at the 1\% level or higher. Errors in the relative flux calibration
  will result in biased flux measurements for large wavelength windows or for wavelength
  windows farther from the \oiii\ emission line.

\item Uncertainties in the continuum subtraction will affect the
  integrated line flux.  Although our Monte Carlo approach accounts
  for part of this uncertainty, there are additional errors introduced
  by the choice of the wavelength windows used to define the
  continuum.  The continuum may also be more complicated than the
  simple linear model employed here.

\item Variable spectral components besides the continuum and
  broad-line emission will affect the light curve measurements.  For
  example, Fe{\sc ii} contamination, which is a problem at all optical
  wavelengths, can add additional variability as it reverberates out
  of phase with both the continuum and the emission lines
  \citep{Barth2013}.  Variable amounts of host-galaxy light also enter
  the spectral extraction aperture due to variations in seeing, which
  will appear as noise in the final light curves \citep{Peterson1995}.
\end{itemize}

Methods besides the rescaling method presented here are necessary to
account for these systematic errors.  Spectral decomposition can help
address the issue of variable spectral components, although such a
decomposition will introduce its own set of uncertainties (i.e.,
model-dependent flux estimates).  Relative flux calibration, on the
other hand, requires great effort and extreme care to reduce below
1\%---even in a recent RM campaign using COS on board of {\it HST}
(for which the sensitivity function is very well known), the
uncertainty floor for repeatability was roughly 1.1\%
\citep{deRosa2015}.

\section{Summary}
We have developed a new procedure for night-to-night calibration of
time-series spectra.  The main innovations of our method are 1) a
common and consistently defined resolution, 2) a more flexible
smoothing kernel, and 3) a Bayesian formalism for fitting the line
profiles and estimating parameter uncertainties.  We have shown that
the method improves the alignment of the [O{\sc iii}] line profiles,
decreasing spurious variability in the rms spectrum and integrated
\oiii\ line light curve.  These improvements help isolate the variable
broad emission-line profiles and reduce night-to-night calibration
uncertainties.  Other systematic effects limit the final precision of
the light curves, such as the calculation of nightly sensitivity
functions and contamination from additional spectral components such
as Fe{\sc ii} emission.  

MMF thanks Richard Pogge for the suggestion to improve the GW92 method,
Chris Kochanek for useful discussions about statistics, and Brad
Peterson for general guidance.  MMF also thanks Kelly Denney, Gisella
de Rosa, Catherine Grier, Kevin Croxall, Richard Pogge, and Brad
Peterson for tips on reducing and analyzing spectroscopic data.  MMF
acknowledges financial support from NSF grant AST-1008882 and a
Presidential Fellowship awarded by The Ohio State University Graduate
School.

\facility{McGraw-Hill}

\software{ Astropy \citep{astropy},
  Matplotlib \citep{matplotlib},
  Numpy \citep{numpy},
  Scipy \citep{scipy} }

\section*{Appendix---Covariances Introduced by the Model}

The rescaling model presented here introduces correlations in the data
because of interpolation and smoothing.  These correlations can affect
integrated quantities, for example, the $\chi^2$ used to fit the model
parameters and the uncertainties in the integrated line flux.  In this
appendix, we assess the importance of these correlations.

The covariance matrix for a vector $\mathbf{\tilde y}$ derived from a vector
$\mathbf{y}$ is
\begin{align}
  {\rm cov}(\tilde y_m,\tilde y_n) = \sum_i\sum_j \frac{\partial \tilde y_m }{\partial
    y_i}\frac{\partial \tilde y_n }{\partial y_j}{\rm cov}(y_i,y_j)
\end{align}
\citep{Gardner2003}.  If $\mathbf{\tilde y}$ is derived from linear
interpolation on $\mathbf{y}$, covariances are introduced between
adjacent points.  Similarly, smoothing introduces local covariances
that depend on the width of the kernel.  The covariances from
smoothing are probably larger than those from interpolation---for
simplicity, we only discuss smoothing in what follows, although the
{\tt mapspec} implementation includes both.

In practice, the convolution in Equation \ref{equ:model} must be
implemented as a discrete sum:
\begin{align}
\tilde O(\lambda_m) &= \sum_i O(\lambda_i) K(\lambda_i - \lambda_m)\\
\tilde O_m &= \sum_i O_i K_m
\end{align}
so the covariance is
\begin{align}
  {\rm cov}(\tilde O_m,\tilde O_n) = \sum_i\sum_j K_mK_n{\rm cov}(O_i,O_j).
\end{align}
Assuming that the original data are uncorrelated, then ${\rm
  cov}(O_i,O_j)$ is diagonal and
\begin{align}
  {\rm cov}(\tilde O_m,\tilde O_n) = \sum_i K_mK_n\sigma_{i}^2,
\end{align}
which reduces to Equation \ref{equ:obserror} if we ignore off-diagonal
terms.  The error spectrum $\sigma_O(\lambda_i) = \sigma_i$ is
estimated during the data reduction/spectral extraction.  It consists
of the photon-counting noise and read noise on each pixel.  In
general, the data in adjacent pixels \emph{are} expected to be
correlated, but the covariance matrix is not known {\it a priori}.  It
may be possible to estimate/model this covariance, perhaps using
Gaussian processes in a manner similar to \citet{Garnett2017}, but
such an analysis is beyond the scope of this work.  We assume that the
original data are uncorrelated here.

\begin{figure}
\includegraphics[width=0.5\textwidth]{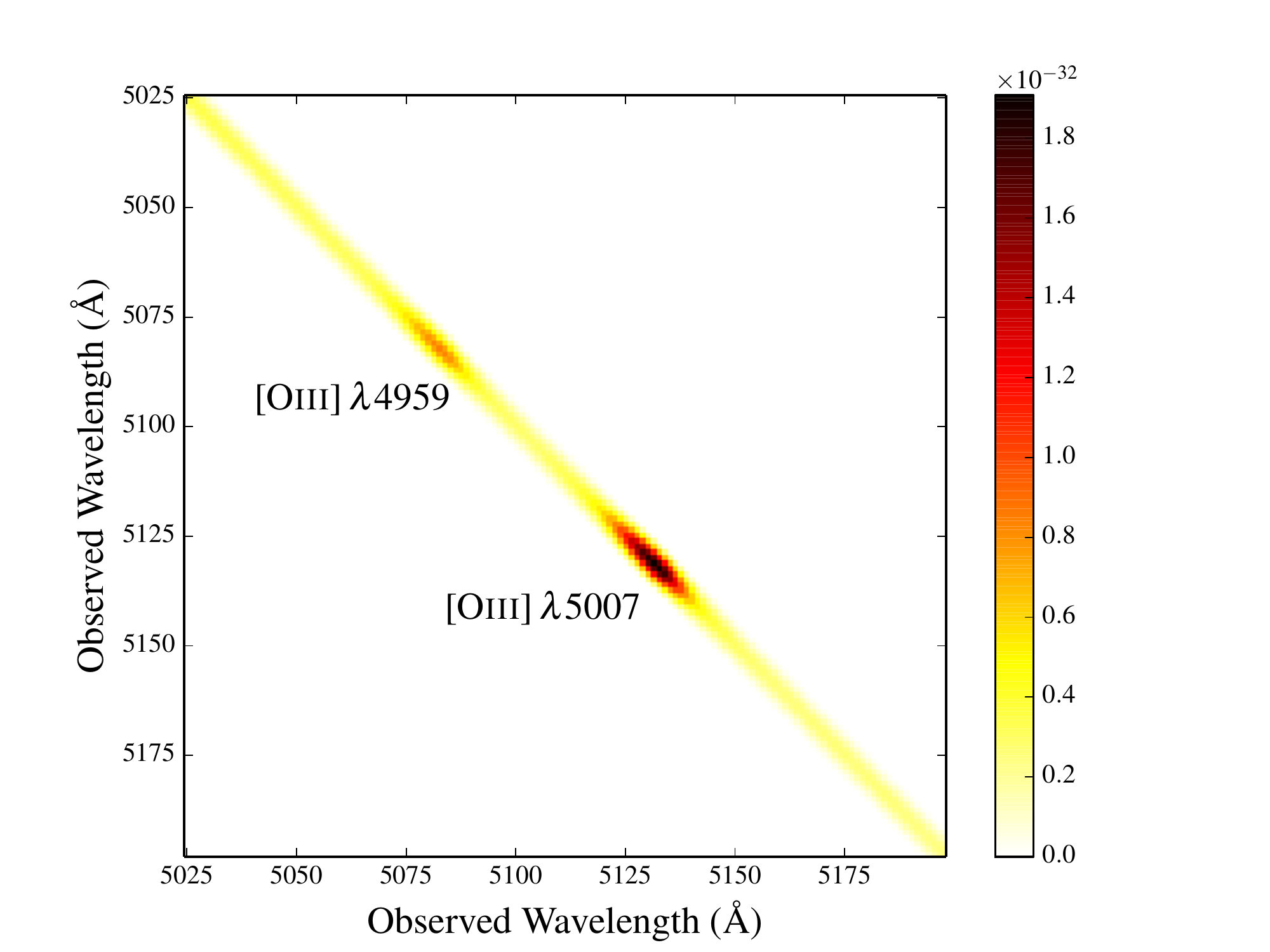}
\caption{Example covariance matrix for wavelengths near the [O{\sc
    iii}] emission lines, calculated for an observation from 2014
  January 10 UT after applying the rescaling model.  The color bar
  shows the covariance in squared physical units, (erg s$^{-1}$
  cm$^{-2}$ \AA$^{-1}$)$^2$.  There are significant local
  correlations, which can affect integrated quantities such as the
  $\chi^2$ used to fit the rescaling model.\label{fig:covar}}
\end{figure}

In Figure \ref{fig:covar}, we show the covariance matrix for
wavelengths near the [O{\sc iii}] lines in the rescaled spectrum from
a photometric night (2014 January 10 UT).  There is clearly a strong
local covariance structure, which is most visible near the emission
lines.  An investigation of the impact of these correlations on the
$\chi^2$ of the {\tt mapspec} fits and the integrated line flux
uncertainty is therefore warranted.

If we define $\mathcal{C} = {\rm cov}(\tilde O_m,\tilde O_n)$, we can
rewrite Equation \ref{equ:loglikely} for a general $\chi^2$
\begin{align}
  \chi^2 = \mathbf{D}^T \mathcal{C}^{-1}\mathbf{D}
\end{align}
where $\mathbf{D} =\hat R(\lambda) - \tilde O(\lambda)$ is the column
vector of residuals.  In principle, $\hat R$ has some associated
covariance matrix $\mathcal{R}$, so we should replace
$\mathcal{C}^{-1}$ with $(\mathcal{C} + \mathcal{R})^{-1}$.  We have
experimented with including this covariance, and found that it makes
virtually no difference to the fits.

\begin{figure}
\includegraphics[width=0.5\textwidth]{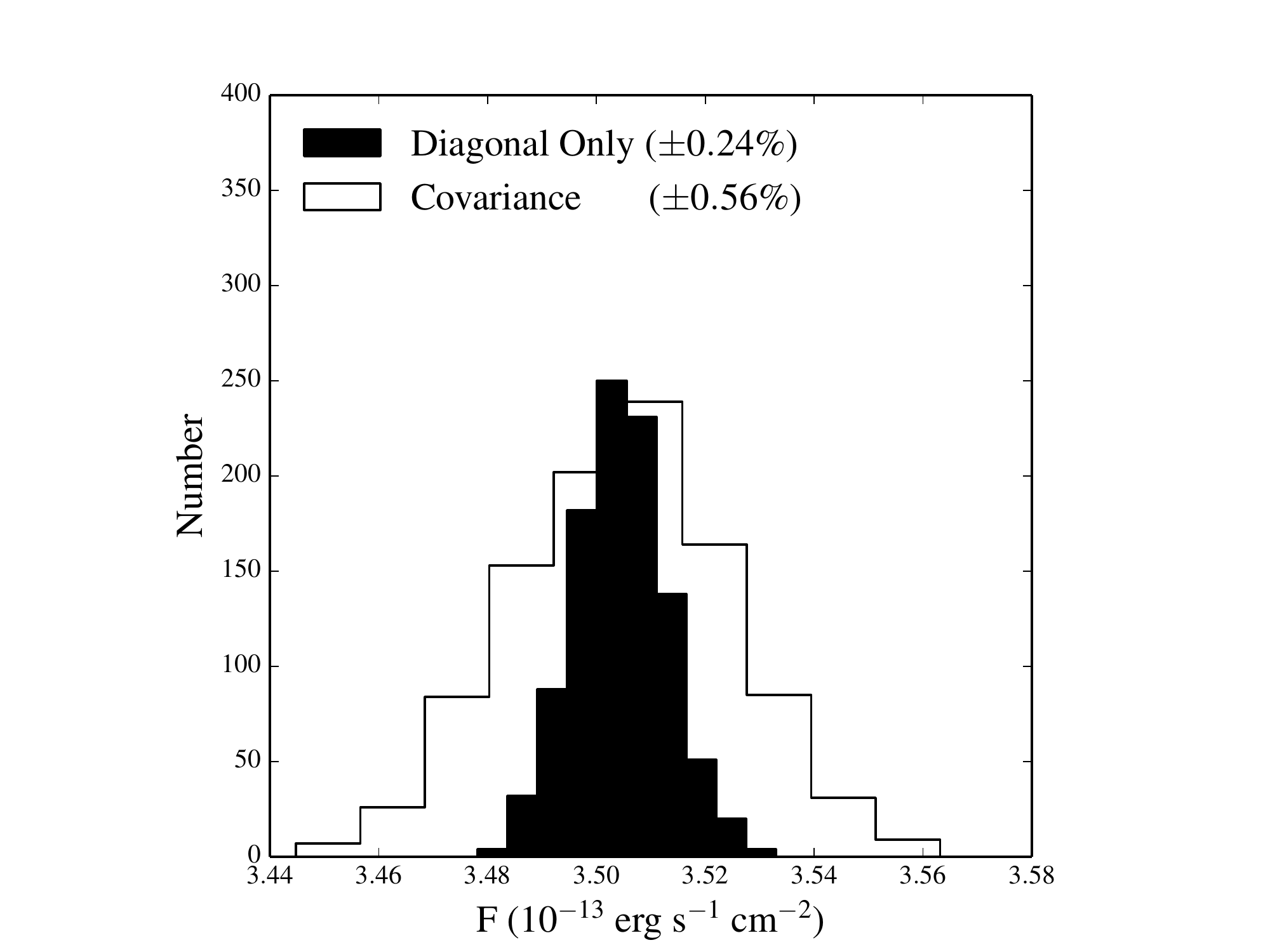}
\caption{Distribution of broad H$\beta$ line fluxes using the Monte
  Carlo methods described in the Appendix.  The widths (central 68\%
  intervals) of the distributions serve as an estimate of the
  integrated line flux uncertainty.  Ignoring correlations in the line
  profile results in underestimated uncertainties by a factor of
  two.\label{fig:covartest}}
\end{figure}

We found that including the full covariance matrix $\mathcal{C}$
reduces the $\chi^2$ by noticeable amounts, but does not change the
best fit parameters of the rescaling model.  This is not
surprising---we can already tell from the good alignment of the \oiii\
line profiles that the fits are nearly optimized (Figure
\ref{fig:spec}).  Including the covariance structure also does not
affect the posterior distributions of the model parameters, although
there is some tendency for the MCMC chains to burn-in faster.
However, this gain does not compensate for the extra time required to
repeatedly calculate and invert $\mathcal{C}$ during the MCMC, so we
revert to Equation \ref{equ:loglikely} when performing the fits.

To quantify the effect of correlations on the uncertainty of the
integrated line flux, we used H$\beta$ as a test case and employed two
methods.  First, we integrated the line profile using a Monte Carlo
method: for $10^3$ iterations, we extracted the line flux after
adjusting the spectrum by random Gaussian deviates scaled to the
measurement uncertainties propagated through the rescaling model.
Second, we repeated this procedure but drew deviates from the
multivariate normal distribution defined by the propagated covariance
matrix.  The first method is equivalent to the second method if we
ignore off-diagonal terms in the covariance matrix.

Figure \ref{fig:covartest} shows the distribution of integrated line
fluxes for each procedure, again using the observation form 2014
January 10 UT.  The presence of correlated errors is extremely
important for the integrated line flux---the width of the distribution
using the covariances is a factor of two larger than that using the
diagonal only.  This result can be intuitively understood through the
integration operation itself---correlated perturbations will tend to
increase/decrease adjacent flux measurements, which magnifies the
change in area under the line profile.  We therefore include the
correlations when calculating the line flux uncertainties in \S3.5.


\end{document}